# Viscoelastic properties of ring-linear DNA blends exhibit non-monotonic dependence on blend composition

Karthik R. Peddireddy[1], Megan Lee[1], Charles M. Schroeder[2], Rae M. Robertson-Anderson[1,*]

[1]Department of Physics and Biophysics, University of San Diego, 5998 Alcala Park, San Diego, CA 92110, United States

[2]Department of Materials Science and Engineering, Beckman Institute for Advanced Science and Technology & Department of Chemical and Biomolecular Engineering, University of Illinois at Urbana-Champaign, Urbana, IL 61801, United States

**ABSTRACT**

Entangled ring polymers, along with blends of ring and linear polymers, continue to be a topic of great interest and debate due to the conflicting experimental results in the literature as well as the difficulty of producing entangled synthetic rings devoid of linear contaminants. Here, we create blended solutions of entangled ring and linear DNA with varying mass fractions of linear DNA $\phi_L$. We use optical tweezers microrheology to measure the linear and nonlinear viscoelastic response of these blends. Our measurements reveal a strong non-monotonic dependence of linear viscoelastic properties on $\phi_L$, with a pronounced maximum when the mass fraction of rings and linear chains are comparable, suggestive of pervasive threading of rings by linear chains. We observe a similar non-monotonicity in the nonlinear regime; however, a comparatively higher fraction of linear chains ($\phi_L \approx 0.5$-$0.7$) is required for a substantial increase in resisitive force and slowing of relaxation dynamics to emerge. This nonlinear response also appears to be rate dependent, which we argue arises from force-induced de-threading of rings at high strain rates. Our results fill a longstanding gap in knowledge regarding the microrheology and nonlinear response of ring-linear polymer blends. Moreover, the uniquely strong mechanical response that ring-linear blends exhibit, along with the ability to finely tune these blends by varying the blend composition, provides new materials design principles.

## I. INTRODUCTION

Ring polymers have been the subject of considerable interest and investigation for the past several decades due to their biological relevance [1-5], industrial applications [6], and intriguing dynamics that are distinct from linear polymer chains [7-17]. However, despite nearly forty years of research on ring polymers, the multitude of conflicting experimental and theoretical results in the literature leaves this issue still of great interest and debate [9,17-19]. The lack of free ends in ring polymers makes understanding their dynamics in the entangled regime particularly challenging [20-22]. In this regime, free ends play an important role in the dynamics of linear polymers, well-described by the reptation model developed by de Gennes and Doi and Edwards [21,23-29]. In this model each entangled linear polymer is allowed to move along its contour, in a "head-first" fashion, but is confined to a tube-like region formed by the surrounding polymers that restricts its transverse motion. While rings have no free ends to undergo traditional reptation, several theoretical models have been proposed to describe the dynamics of entangled ring polymers [7,18,19,30,31]. Possible diffusive mechanisms that have been proposed include: modified reptation, in which the ring assumes a folded conformation akin to a linear chain of half its length; mutual penetration, in which the rings thread each other; and amoeba-like motion, in which the ring forms multiple arms that it uses to explore its surroundings [7,18,30,32-37].

The longest relaxation time for entangled rings, i.e. the disengagement time, has been predicted and experimentally shown to be shorter than that for linear chains [32,38]. At the same time, the extent to which rings can even form entanglements remains a topic of debate [7,15,21,39,40]. For example, a recent report shows that linear viscoelastic properties of entangled ring polymers are better described by the Rouse model for unentangled polymers instead of the above mentioned mechanisms [15]. Further, both entangled linear and ring polymers have been shown to exhibit strain hardening, in which the stiffness of the material increases as strain is increased [8,41-43]. However, despite their faster relaxation times, ring polymers exhibit more pronounced hardening that occurs at larger strains and persists for longer times than their linear counterparts [41,44]. In fact, we recently reported evidence of strain hardening even in semidilute unentangled blends of ring and supercoiled DNA [45].

Much of the conflicting reports for entangled rings have been attributed to the varying degree to which linear polymer 'contaminants' are present in synthetic ring samples [9,18,19,46,47]. While considerable effort has been made over the past few decades to improve the cyclization process used to synthesize ring polymers, entangled systems of 100% pure synthetic rings have yet to be achieved [19,48,49]. This issue is complicated by the fact that even a small fraction of linear polymers has been shown to have a profound effect on the dynamics of synthetic ring polymers, leading to increased viscosity, hindered diffusion and rubbery plateaus that are absent in nearly pure ring systems [7,9,18,47,50-54]. This extreme sensitivity to linear polymers has been postulated to arise from linear chains threading rings and essentially halting their center-of-mass motion [18,50,55]. The only mechanism whereby threaded rings can diffuse or relax is via constraint release of the threading linear chains – an extremely slow process compared to reptation. Several other ring-linear entanglement mechanisms such as once-threaded [52] and unthreaded-linear [14] models have also been proposed. Yet, the role that each mechanism plays in the viscoelastic response of entangled ring-linear blends is still debated [14,54,56,57].

Most of the studies to date on ring-linear blends have focused on steady-state dynamics and unentangled or marginally entangled systems, and have reported conflicting results [9,16,18,19,21,46,50,55,58]. For example, the viscosity of ring polymer systems has been shown to increase with the addition of linear polymers, reaching values >2x greater than that of pure linear polymers. However the exact dependence of viscosity on linear polymer mass fraction ($\phi_L$) is not yet settled [16,18,19,46,55,58]. Further, very few studies have examined the response of ring-linear blends to nonlinear strains [15,53].

Previously, we investigated the diffusive behavior of DNA in blended solutions of entangled ring and linear DNA with varying $\phi_L$ [50]. We showed that as $\phi_L$ increased the diffusion coefficient of rings sharply dropped, until $\phi_L=0.5$, after which it maintained a $\phi$-independent plateau with values that were lower than their linear counterparts. Conversely, the diffusion coefficient for linear DNA displayed a non-monotonic dependence on $\phi_L$, reaching a minimum at $\phi_L=0.5$. Our corresponding simulations showed that this surprising non-monotonic behavior arose from second order effect of threading of rings by linear chains. Namely, at $\phi_L=0.5$, every ring can be threaded by a linear molecule, as there are an equal number of rings and linear chains, so the system can effectively be comprised entirely of threaded rings. The highly restricted motion of threaded rings leads to the most restrictive environment for entangled linear polymers to diffuse through. As $\phi_L$ increases beyond 0.5, threaded rings are replaced with entangled linear chains, which are more mobile than threaded rings as they are free to reptate. Similarly, as $\phi_L$ decreases below 0.5, the number of threading events is reduced as linear chains that were threading rings are replaced with rings which are much less effective at threading, leading to a more mobile system.

Here, to build on these steady-state results and to determine the robustness of ring-linear entanglements and threading to nonlinear straining, we perform linear and nonlinear optical tweezers microrheology on entangled blends of 45 kbp (15 μm) ring and linear DNA. We find that ring-linear DNA blends exhibit a strong non-monotonic dependence on $\phi_L$, with blends with intermediate $\phi_L$ values (~0.3-0.7) exhibiting the highest rubbery plateau and the most pronounced shear-thinning in the linear regime. In the nonlinear regime, these blends display the largest resistive force, the highest effective viscosity, and the slowest relaxation dynamics. However, the variation between blends is distinct in the nonlinear regime compared to the linear regime, suggestive of forced de-threading. Our suite of results demonstrates that threading of rings by linear chains is indeed most pervasive with comparable fractions of ring and linear chains, and plays a principle role in the dynamics of ring-linear blends.

## II. MATERIALS AND METHODS

### A. Sample Preparation

Double-stranded DNA molecules with a contour length of 45 kilobasepairs (kbp) were prepared by replication of cloned fosmid constructs in Escherichia coli, followed by extraction, purification and concentration, as detailed thoroughly elsewhere [59,60]. Following this process, the 2.4 mg/ml DNA solution was comprised of 20% supercoiled circular DNA, 66% relaxed circular (ring) DNA, and 14% linear DNA in TE10 buffer (10 mM Tris-HCl (pH 8), 1 mM EDTA, and 10 mM NaCl). The concentration and percentages of each topology were determined via agarose gel electrophoresis and single-molecule flow experiments as described previously [45,53]. To prepare 100% linear DNA, a fraction of the solution was treated with the restriction enzyme ApaI. To prepare 14% linear (86% ring) DNA samples, the remaining fraction of the solution was treated with Topoisomerase I to relax supercoils. 32%, 50%, 68% and 86% linear DNA blends were prepared by mixing of the two stock solutions. All samples for experiments were diluted to 1 mg/ml in TE10. Our previous measurements examining the concentration dependence of the diffusion coefficients for the 45 kbp ring and linear DNA used here showed that the critical entanglement concentration ($c_e$), determined as the concentration at which the scaling shifts from Rouse to reptation scaling, is ~0.3 mg/ml for both ring and linear topologies [61]. The corresponding number of entanglements per chain $N_e$, determined via the relation $N_e = (c/c_e)^{1.25}$ established for linear polymers, is ~4-5 [62,63].

For microrheology experiments, a trace amount of 4.5 μm polystyrene microspheres, coated with Alexa-488 BSA to prevent DNA adsorption and enable fluorescence visualization, were added to the DNA solutions. 0.1% Tween-20 was also added to the solution to prevent adsorption of DNA to the sample chamber walls. As such, the boundaries between the polymers and the beads and surfaces can be considered no-stick boundaries. Further, to ensure that we are probing the rheology of the DNA network rather than the non-continuum local rheology, we chose a bead radius that was >3x the entanglement tube radius $a$ of our networks ($a_L \approx 0.27$ μm and $a_R \approx 0.22$ μm for pure linear and ring DNA solutions, respectively) [8,32,38,45,60]. This criterion has been theoretically and empirically shown to be sufficient to probe the continuum mechanics of entangled polymer solutions [64-67]. Nonetheless, it has also been shown that microrheology typically underestimates the magnitudes of $G'$ and $G''$ compared to bulk rheology; however, their dependences on frequency and sample concentration are transferable between the two techniques [68-71]. As such, to facilitate comparison to bulk rheology we focus our discussion on the scalings and trends in the data rather than the absolute magnitudes.

Samples were mixed slowly using wide-bore pipette tips to prevent shearing and breaking of rings. The samples were then further allowed to equilibrate by slow rotation (8 rpm) for at least 30 minutes. A sample chamber was made with a microscope glass slide, a cover slip and two small pieces of double-stick tape as spacers. The chamber was filled with the DNA solution through capillary action and then hermetically sealed with epoxy and allowed to equilibrate for a minimum of 15 minutes before measurements.

### B. Microrheology

We used optical tweezers microrheology to determine the linear and nonlinear dynamics of ring-linear blends (Fig. 1b-f). Details of the experimental procedures and data analysis, briefly summarized below, have been described in detail in refs [72,73]. The optical trap consisted of an Olympus IX7I microscope with a 60x 1.4 NA objective (Olympus) and a 1064 nm Nd:YAG fiber laser (Manlight). A position sensing detector (Pacific Silicon Sensors) measured the deflection of the laser beam, which is proportional to the force exerted by the solution on the trapped bead. The proportionality constant (i.e. trap stiffness) was obtained using Stokes drag method.

Linear viscoelastic properties were determined from thermal fluctuations of a trapped microsphere, measured by recording the associated laser deflections for 100 seconds. Linear viscoelastic moduli, i.e. the elastic modulus $G'(\omega)$ and the viscous modulus $G''(\omega)$, were extracted from the thermal fluctuations using the generalized Stokes-Einstein relation as described in ref [74]. The procedure requires the extraction of normalized mean-squared displacements ($\pi(\tau) = <r^2(\tau)>/2<r^2>$) of the thermal forces, averaged over all trials, which is then converted into the Fourier domain via:

$$-\omega^2 \pi(\omega) = \left(1 - e^{-i\omega\tau_1}\right)\frac{\pi(\tau_1)}{\tau_1} + \dot{\pi}_\infty e^{-i\omega t_N} + \sum_{k=2}^{N}\left(\frac{\pi_k - \pi_{k-1}}{\tau_k - \tau_{k-1}}\right)\left(e^{-i\omega\tau_{k-1}} - e^{-i\omega\tau_k}\right),$$

where $\tau$, 1 and N represent the lag time and the first and last point of the oversampled $\pi(\tau)$. $\dot{\pi}_\infty$ is the extrapolated slope of $\pi(\tau)$ at infinity. Oversampling is done using the MATLAB function PCHIP. $\pi(\omega)$ is related to viscoelastic moduli via:

$$G^*(\omega) = G'(\omega) + iG''(\omega) = \left(\frac{k}{6\pi R}\right)\left(\frac{1}{i\omega\pi(\omega)} - 1\right),$$

where $R$ and $k$ represent the radius of the microsphere and trap stiffness.

We computed the complex viscosity $\eta^*(\omega)$ via $\eta^*(\omega) = [(G'(\omega))^2 + (G''(\omega))^2]^{1/2}/\omega$. We further converted $G'(\omega)$ into the stress relaxation modulus $G(t)$ via:

$$G(t) = 2/\pi \int_0^\infty (G'/\omega) \sin\omega t \, d\omega.$$

In practice, we obtained $G(t)$ using $G'(\omega)$ for the range of frequencies available. Numerical integration was done using the TRAPZ function in MATLAB.

The Doi-Edwards (D-E) model [8] predicts viscoelastic properties of entangled linear polymers and the predicted elastic modulus is given by the equation

$$G'(\omega) = G^0 \sum_{p;\, odd} \frac{8}{\pi^2} \frac{1}{p^2} \frac{(\omega\tau_{D,L}/p^2)^2}{(1+(\omega\tau_{D,L}/p^2)^2)}, \text{ for } \omega\tau_{e,L} \leq 1,$$

where $G^0$, $\tau_{D,L}$ and $\tau_{e,L}$ are the elastic plateau modulus, disengagement time and the relaxation time for an entangled linear polymer. $\tau_{e,L} = a_L^4/24R_{G,L}^2 D_L$ and $\tau_{D,L} = 36R_{G,L}^4/\pi^2 a_L^2 D_L$ where $a_L$ is the entanglement tube radius, $R_{G,L}$ is the radius of gyration, and $D_L$ is diffusion coefficient in dilute conditions. D-E model further relates $G^0$ to the number of entanglements per chain as $N_e = (4/5)ck_B N_A T/(MG^0)$ where $k_B$, $N_A$, $T$ and $M$ are Boltzmann constant, Avogadro's number, temperature and molecular weight of the polymer. The relation between $a_L$ and $N_e$ is given by $a_L = (24\,N_e/5)^{1/2}\,R_{G,L}$ [8].

Nonlinear microrheology measurements were performed by displacing a trapped microsphere through the sample at speeds of $v = 10 - 80$ μm/s using a piezoelectric nanopositioning stage (Mad City Laboratories) to move the sample relative to the microsphere. Speeds were converted to strain rates via $\dot{\gamma} = 3v/\sqrt{2}R$ (9.4-75s$^{-1}$) [75]. While these types of microrheological strains are typically assumed to be more analogous to shearing rather than extensional bulk rheology, because we are pulling a microsphere through the blends, there may be components of extensional rheology at work as well, as DNA strands can get momentarily hooked on the bead before slipping off [45].

For both linear and nonlinear measurements, all data was recorded at 20 kHz and at least 15 trials were conducted, each with a new microsphere in an unperturbed location. Presented data is an average of all trials.

## III. RESULTS AND DISCUSSION

To characterize the linear regime rheological properties of the blends, we extract the frequency-dependent elastic modulus, $G'(\omega)$ and complex viscosity $\eta^*(\omega)$ from the thermal fluctuations of trapped beads (Figs. 1b,e and 2). As shown in Fig. 2a, the elastic moduli for all blends show similar frequency dependence, increasing with frequency at low frequencies then approaching a frequency-independent elastic plateau, $G^0$. However, the magnitudes of $G'(\omega)$ and $G^0$ display a non-monotonic dependence on $\phi_L$, with a ~4x increase from $\phi_L=0.14$ to $\phi_L=0.32$, followed by a further more modest increase to 0.68, followed finally by a ~3-fold drop to $\phi_L=0.86$ and 1 blends. (Fig. 2a,d). To verify our data and determine the role of the entangled rings versus linear chains in the blends, we compare our data to known theories for entangled linear and ring polymers [8,18,19]. The Doi-Edwards model for entangled linear polymers predicts $\tau_D = 36R_G^4/\pi^2 a_L^2 D$ which we calculate as $\tau_D \cong 1.8$s using our measured values for $R_G$ and $D$ [60]. Using this value for $\tau_D$, we find reasonable agreement between our experimental $G'(\omega)$ curve for pure linear DNA and the D-E model predictions (described in Methods). Minor discrepancies between our data and the D-E model may be due to the relatively low density of entanglements in our systems ($N_e \cong 4$). In this regime, tube length fluctuations and constraint release may also contribute to the response. Two well-known models for entangled rings, the lattice-animal (L-A) model and the Rouse model, are also shown. The lattice-animal model [18] predicts abnormally high values for pure rings, as the predicted $G'(\omega)$ curve is higher than the experimental values observed for the $\phi_L=0.14$ DNA blend. Because this model is intended for pure ring systems, and the presence of linear contaminants is expected to increase the modulus, we should expect the model curve to be below the experimental curve for $\phi_L=0.14$ if the model accurately captures the dynamics of entangled rings. On the other hand, the $G'(\omega)$ curve predicted by the Rouse model is consistently below the experimental curve for $\phi_L=0.14$ DNA blend. A few other recent studies have observed similar trends and inconsistencies with predictions for ring-linear blends, suggesting that refinement of the theories for entangled rings and ring-linear blends is needed [15,19].

We convert our $G'(\omega)$ curves to time-dependent stress relaxation moduli $G(t)$ as described in the methods section (Fig. 2b). From $G(t)$ we can also estimate an elastic plateau value $G_N$ from the $G(t)$ value at the shortest measured timescale, where the data is approaching a time-independent plateau. Stress relaxation curves for all blends exhibit multimodal exponential decay, as is expected for entangled linear polymers [8,76]. However, the magnitudes of the $G(t)$ curves and $G_N$ values show a significant non-monotonic dependence on $\phi_L$, similar to that for $G'(\omega)$ and $G^0$, suggesting a strong influence of topology on the equilibrium stress growth with time (Fig. 2b,d). The reptation model predicts $G_N \sim c^2$ where $c$ is the total polymer concentration. Because we fix $c$ to 1 mg/ml for all blends, if topology did not significantly influence dynamics then we should expect all elastic plateau values to match. Conversely, we measure a strong non-monotonic dependence on $\phi_L$ indicating that the nature of the entanglements is changing for the different blends. Past studies on synthetic ring-linear blends have shown similar topology dependence of $G^0$, with a plateau becoming increasingly more apparent and of higher magnitude as $\phi_L$ increases from 0 to 0.2 [18,46]. Similar to our measurements $G^0$ values for blends with $\phi_L<0.2$ remain significantly lower than for the pure linear system ($\phi_L= 1$). Another very recent study on synthetic ring-linear blends with $\phi_L = 0.7$-1 have shown that the stress relaxation curves increase as $\phi_L$ decreases from 1 to 0.7, in line with our results [77]. Finally, our result for the $\phi_L= 0.86$ DNA blend aligns with a recent report for a synthetic ring-linear polymer blend with $\phi_L= 0.85$ [15]. In our work and that of Ref 15, the $G_N$ value for the blend is nearly identical to that for $\phi_L= 1$. No previous studies to our knowledge have reported plateau values for ring-linear blends with $0.2 <\phi_L<0.7$.

We next evaluate the frequency-dependence of the complex viscosity $\eta^*(\omega)$ for our blends (Fig. 2c,d). According to the Cox-Merz rule, which has been shown to be valid in the linear regime for both entangled rings and linear polymers (including DNA) [15,64,78,79], the complex viscosity from oscillatory shear measurements ($\eta^*(\omega)$) and the dynamic viscosity from steady-shear experiments ($\eta(\dot{\gamma})$) can be used interchangeably [80]. As a result, we can compare the frequency-dependence of our measured $\eta^*(\omega)$ curves to predictions and previous reports for the rate-dependence of $\eta(\dot{\gamma})$. As shown, all measured $\eta^*(\omega)$ curves exhibit shear thinning, in which the viscosity decreases with increasing strain rate according to the power-law $\eta^* \sim \omega^{-\alpha}$. As shown in Fig. 2d, the scaling exponents display the signature non-monotonic dependence on $\phi_L$, with the $\phi_L=0.14$ blend exhibiting the weakest thinning while the maximum exponent is reached for $\phi_L=0.5$-0.68. Further, the thinning exponent of ~0.6 for $\phi_L=1$ matches with our previously measured exponent for entangled linear DNA [64] as well as simulation results based on the finitely extensible nonlinear elastic chain model for polymer melts [81]. It has been previously shown that entangled ring polymers exhibit weaker shear thinning in comparison to their linear counterparts, in line with our results for $\phi_L=0.14$ and $\phi_L=0.86$ [15]. In this previous work, weaker shear thinning was hypothesized to be a result of the inability of rings to deform and stretch in the direction of strain as easily as linear chains. Enhanced shear thinning in blends with $\phi_L=0.32$-0.68, in comparison to the $\phi_L=1$ blend, suggests that pervasive threading of rings by linear DNA helps ring DNA to align in the direction of strain.

Our collective linear microrheology results reveal that the DNA blends with $\phi_L=0.5$ and $\phi_L=0.68$ have the strongest spatial constraints while the $\phi_L=0.14$ DNA blend has the weakest constraints. To determine the robustness of entanglements and threading to large strains, we turn to our nonlinear microrheological measurements (Figs. 1b-c,f, 3 and S1). As described in Methods, to characterize the nonlinear response of the blends, we optically drive a 4.5 μm microsphere 30 μm through the blends at strain rates of $\dot{\gamma}=9.4$–75s$^{-1}$ (Fig. 3). We chose the distance and rates to ensure we are probing the nonlinear regime. For reference, the strain distance equates to a strain ($\gamma$) of 6.7 which is much higher than the critical value of 1 for nonlinearity [13]. Another necessary and sufficient condition for the nonlinear regime is that strain rates must be higher than a certain terminal relaxation frequency, $\omega_T = \lim_{\omega \to 0} \omega G''/G'$ [13]. As shown in Fig. S2, while our data does not exactly reach the terminal relaxation regime, the lowest frequency values provide an upper-bound of ~0.1 s$^{-1}$ for $\omega_T$. The strain rates we use are clearly higher than $\omega_T$ for all blends. To further elucidate the nonlinear nature of our force curves, we compute stress curves from $G(t)$ using $\sigma_{LVE}(t) = \dot{\gamma} \int_0^t G(t)dt$ and compare them with our measured nonlinear stress curves (Figs. 2b, S3) [13]. Stresses curves for linear and nonlinear microrheology techniques differ by an order of magnitude (Fig. S3). The stress is maximum for the $\phi_L=0.68$ DNA blend in both techniques but the exact dependence of values on $\phi_L$ is substantially different in the nonlinear regime. Notably, in the nonlinear regime, the $\phi_L$=0.32 blend is nearly identical to $\phi_L$=0.14 while it is substantially larger in the linear regime. These data demonstrate that we are indeed probing two entirely different regimes in our linear and nonlinear measurements.

As shown in Figs. 3a and S1, the nonlinear stress curves for all blends initially rise steeply before reaching a 'softer' regime in which the slopes of the force curves are shallower. Further, as better shown in Fig. 3b, in which the force is plotted on a log-scale versus time, following initial softening all blends subject to high strain rates exhibit a strain-hardening regime in which the slopes of the force curves increase. As mentioned in the Introduction, a recent report on extensional rheology of linear and ring polystyrene shows that rings have a significant delayed strain hardening response in comparison to their linear counterpart at all extensional strain rates [41]. However, we observe no such delays in our ring-linear blends which suggests that a small linear fraction is sufficient to effectively guide the stretching of ring polymers.

While the strain dependence of the force response has similar features for all blends, the magnitude of the force response follows a non-monotonic dependence on $\phi_L$. To better evaluate this non-monotonicity we plot the maximum force reached during strain versus linear DNA fraction (Fig. 3c) and strain rate (Fig. 3d). As shown, the $\phi_L$=0.14 and 0.32 blends produce the weakest response at all strain rates while the $\phi_L$=0.68 blend exhibits the strongest. The $\phi_L$=0.5, 0.86 and 1 blends elicit forces in between these two extremes. Further, Fig. 3d shows that the maximum force for all blends exhibits a linear dependence on strain rate ($F_{max} \sim \dot{\gamma}$) with slopes that depend on $\phi_L$. From the slope of each $F_{max}(\dot{\gamma})$ curve we can approximate an effective viscosity $\eta_{eff}$ using Stokes law $F_{max} = 6\pi\eta_{eff}Rv$. As shown in Fig. 3e, the effective viscosity displays the same non-monotonic dependence on $\phi_L$ as our other metrics.

Finally, we evaluate the time at which each force curve initially 'softens' or transitions to a weaker strain dependence, which we term the softening time $t_{soft}$. As shown in Fig. 3f, for all blends $t_{soft}$ generally decreases with increasing strain rate, converging to a rate-independent value very close to the theoretically predicted Rouse relaxation time for linear polymers ($\tau_{R,L} = 6R_{G,L}^2/3\pi^2 D_L \approx 0.11$s) but significantly higher than that for ring polymers ($\tau_{R,R} \approx 0.04$ s). Interestingly we observed a very similar trend at high strain rates for semidilute blends of ring and supercoiled DNA [45]. In this previous work, we found that at high strain rates, $t_{soft}$ values converged to the theoretically predicted Rouse time for ring DNA ($\tau_{R,R}$) with no apparent contributions from the supercoiled constructs. We rationalized this result as arising from the nonlinear strain forcing the separation of the rings and supercoiled molecules. The moving probe forced the faster, more compact supercoiled molecules to disentangle from the rings and sweep past the probe into its wake. At the same time, the slower, more extended rings built up in front of the moving probe and thus dictated the measured force relaxation. Likewise, the convergence of $t_{soft}$ values for all blends to $\sim\tau_{R,L}$ may arise from force-induced separation of rings and linear chains, with linear DNA building up in front of the probe and rings de-threading and disentangling from linear DNA and falling behind the moving probe.

Following the applied nonlinear strain, the microsphere is halted and the relaxation of the force is measured over time (Figs. 1d,f, 4). As shown in Fig. 4a, all blends relax to equilibrium conditions (i.e. $F_{final} = 0$) although with varying relaxation rates. We extract relaxation timescales by fitting each curve with a triple exponential decay function ($F(t)=C_1 e^{-t/\tau 1}+C_2 e^{-t/\tau 2}+C_3 e^{-t/\tau 3}$). We have previously shown that this function can describe the relaxation dynamics of entangled linear and ring DNA as well as semidilute ring-supercoiled DNA blends [38,45,82]. This function fits our data well with adjusted R-squared values of 0.99 and higher and with three distinctly different time constants (Fig. 4b). Single or double exponentials do not fit the data well and fits do not converge when we add more exponential decay terms. We find that the measured time constants are independent of strain rate but they do depend on $\phi_L$ (albeit weakly). Fig. 4b shows the measured time constants, averaged over all strain rates, for each $\phi_L$. As shown, the $\phi_L$=0.14 blend has the fastest relaxation timescales and the $\phi_L$=0.68 blend has the slowest, as expected from the data shown in Figs. 2 and 3.

To determine the relaxation mechanisms responsible for the three distinct relaxation timescales, we compare our measured time constants to the three principle relaxation timescales predicted by the reptation model for entangled linear polymers. The fastest timescale is the entanglement time $\tau_{e,L}$ which is the timescale over which thermally diffusing chain segments reach the edge of the reptation tube. The slowest timescale, the disengagement time $\tau_{D,L}$, is the time over which the polymer reptates completely out of its initial deformed tube. The intermediate timescale is the Rouse time $\tau_{R,L}$, or the time over which elastic relaxation of the deformed polymer occurs. Within this framework, the predicted timescales for our linear DNA solution ($\phi_L$=1) are $\tau_{e,L} \cong 0.03$ s, $\tau_{R,L} \cong 0.11$ s, and $\tau_{D,L} \cong 1.8$ s [8,60]. As shown in Fig. 4b, our three measured time constants ($\tau_1, \tau_2, \tau_3$) are comparable to these predicted times for all DNA blends. Conversely, the predicted timescales for a pure ring solution, based on the pom-pom ring model, are $\tau_{e,R} \cong 0.007$ s, $\tau_{R,R} \cong 0.04$ s, and $\tau_{D,R} \cong 0.17$ s [32,45]. These quantities specifically come from the following relations predicted by this model: $(a_R/a_L)^2 \sim (L/2p)^{-1/2}$ and $\tau_{D,R}/\tau_{D,L} = (a_R/a_L)^2 (L/2p)^{-1/2}$ where $p$ is persistence length [32]. This result corroborates the high strain rate softening time result which also aligns with the Rouse time for linear DNA for all blends.

Therefore, it appears that strong nonlinear forcing can indeed disrupt ring-linear entanglements and/or de-thread rings. As described above, because rings have faster relaxation dynamics compared to linear chains and cannot as easily stretch in the direction of strain, rings are able to more easily sweep past the moving probe than linear chains. The result is that entangled linear DNA builds up in front of the probe while rings are left in its wake. However, due to the pervasive threading in the $\phi_L$=0.50 and $\phi_L$=0.68 blends we do not expect these systems to become completely de-threaded so we expect that the relaxation timescales should exhibit a similar non-monotonic dependence as our other metrics that appear to be controlled by threading.

To further test this interpretation and elucidate the relaxation dynamics of the blends, we evaluate the fractional coefficients $C_i$ associated with each decay mode. Unlike the decay times, the coefficients exhibit little dependence on $\phi_L$ but significant dependence on strain rate. As such we evaluate the average $C_i$ value across all blends as a function of strain rate (Fig. 4c). As shown in Fig. 4c, while the degree to which blends undergo Rouse relaxation (i.e. $C_2$) is relatively insensitive to strain rate, $C_3$ decreases with strain rate while $C_1$ increases. As such it appears that for larger strains the effect of threading and entanglements weakens, likely by the forced separation of rings and linear chains. The non-monotonic dependence we see for our relaxation times further corroborates this interpretation. We would only expect the non-monotonic dependence of $\tau_3$ to persist if the system remained highly threaded such that constraint release was contributing to $\tau_3$ to make it slower. If fast strains can effectively de-thread rings from linear chains, the degree to which blends undergo the slow mode should decrease with increasing strain rate, which is in fact what we see in Fig. 4c. Thus, threading indeed has a subdued effect at these fast modes.

## IV.    CONCLUSIONS

Here, we present linear and nonlinear optical tweezers microrheology measurements of entangled blends of ring and linear DNA. We observe a strong non-monotonic dependence of linear viscoelastic properties on $\phi_L$, with a pronounced maximum when the mass fraction of rings and linear chains are comparable. We argue that this non-monotonicity is a result of threading of ring polymers by linear chains coupled with the relative ineffectiveness of rings to self-entangle compared to linear polymers. Pervasive threading in the $\phi_L$=0.68 blend leads to a higher elastic plateau value as well as more pronounced shear-thinning compared to the pure linear system ($\phi_L$=1).

Our nonlinear microrheology results reveal that ring-linear threading is robust to modest nonlinear strains but can be disrupted at very high strain rates (>50 s$^{-1}$). This force-induced de-threading causes entangled linear DNA to build up in front of the moving probe while rings, which are less effective at stretching and orienting in the direction of the strain, slide past the probe into the wake. This process results in the linear polymers playing the principle role in the nonlinear relaxation dynamics of ring-linear blends.

Our results provide important new insights into the dynamics of entangled ring polymers and ring-linear blends – topics of current interest and debate. In particular, we have addressed the dearth of experimental data on the microrheology as well as the nonlinear response of ring-linear polymer blends. As such, we anticipate that our work - which highlights the importance of dynamic threading events to the rheology of ring-linear blends – will prompt new theoretical investigations of the response of topological polymer blends across wide-ranging spatiotemporal scales. Finally, the emergent strong viscoelastic response that ring-linear blends exhibit, along with the ability to finely tune the rheological properties of these blends by varying the relative fractions of each topology, suggest important potential industrial applications.

## FOOTNOTES

**Supporting Information.** Figure S1. Nonlinear stress responses of ring-linear DNA blends; Figure S2. Terminal relaxation frequency and time as determined from linear microrheology experiments; Figure S3. Comparison between expected linear viscoelastic elastic (LVE) stress growth and measured nonlinear stress growth in nonlinear microrheology experiments.

## AUTHOR INFORMATION

**Corresponding Author**

*Email: randerson@sandiego.edu

**ORCID**

Rae M. Robertson-Anderson: 0000-0003-4475-4667

## CONFLICTS OF INTEREST

There are no conflicts to declare.

## ACKNOWLEDGEMENTS

The authors acknowledge financial support from Air Force Office of Scientific Research (AFOSR-FA9550-17-1-0249) and National Science Foundation (NSF-CBET-1603925).

# FIGURES

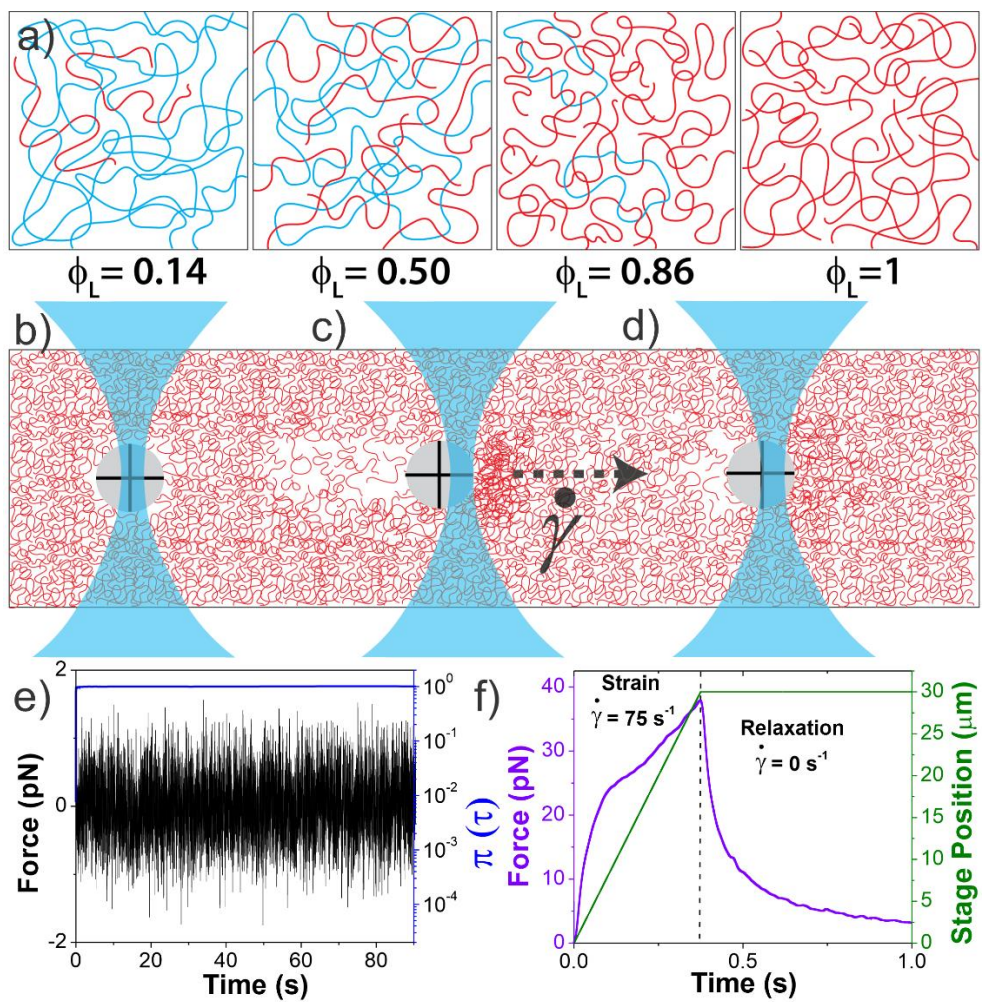

**Figure 1. Experimental approach to probe the rheological properties of entangled ring-linear blends.** (a) Cartoon of blends of ring (blue) and linear (red) DNA of identical contour lengths at four different mass fractions of linear DNA $\phi_L$. (b-d) Cartoon of optical tweezers microrheology with polymer sizes increased ~4$x$ for better visibility. (b) A 4.5-μm microsphere (grey circle) embedded in the DNA solution and trapped using a focused Gaussian laser beam (blue). At equilibrium, centers of the bead and beam are, on average, perfectly aligned. Linear microrheology measurements are performed by measuring the thermal deviations of the bead from the trap center in equilibrium. (c) The same optically trapped 4.5-μm bead is displaced 30 μm through each blend at speeds $v$ = 10–80 μm/s, corresponding to strain rates $\dot{\gamma} = 3v/\sqrt{2}R = 9.4 - 75$ s$^{-1}$ where $R$ is the bead radius. The bead center is displaced from the laser center due to the force exerted by the surrounding polymers when the particle is dragged through the solution. (d) The bead motion is halted and the surrounding polymers relax back to equilibrium, allowing the bead to return to the center of the trap. (e) Linear microrheology. An example of thermal oscillation data for the $\phi_L$=0.14 DNA blend. The data is captured for 100 seconds at 20 kHz. We extract normalized mean square displacements ($\pi(\tau)$) from the thermal oscillations which then are used to extract viscoelastic moduli using the generalized Stokes-Einstein relation (see Methods). (f) Nonlinear microrheology. Stage position (green) and force exerted on the trapped bead (violet) during (0.4-6 s), and following (9-15 s) the bead displacement (delineated by dashed lines) are recorded at 20 kHz. Data shown is for the $\phi_L$=0.14 DNA blend at $v$ = 80 μm/s.

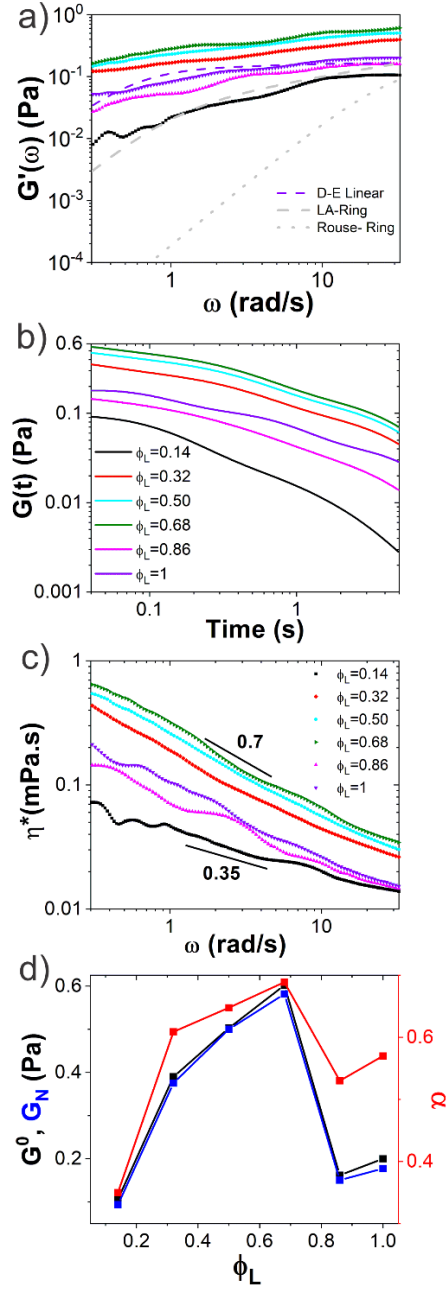

**Figure 2. Linear frequency-dependent viscoelastic moduli of ring-linear blends exhibit strong non-monotonic dependence on linear DNA fraction** $\phi_L$. (a) Frequency-dependent elastic modulus $G'(\omega)$ for varying linear DNA mass fractions $\phi_L$. All $G'(\omega)$ curves approach elastic plateaus at high frequencies with a non-monotonic dependence of the magnitude on $\phi_L$. Theoretical curves predicted from the Doi-Edwards (D-E) model for linear polymers [8], the lattice animal (L-A) model for rings [18], and the Rouse model for rings [15] are plotted for comparison. (b) Time-dependent stress relaxation modulus $G(t)$ for varying $\phi_L$. (c) Complex viscosity $\eta^*(\omega)$, showing varying degrees of shear thinning ($\eta^* \sim \omega^{-\alpha}$) with non-monotonic dependence on $\phi_L$. (d) Elastic plateau modulus ($G^0$, black) determined from $G'(\omega)$, initial relaxation modulus ($G_N$, blue) determined from $G(t)$, and shear thinning exponent ($\alpha$, red) determined from power-law fits to $\eta^*(\omega)$, all plotted as a function of $\phi_L$.

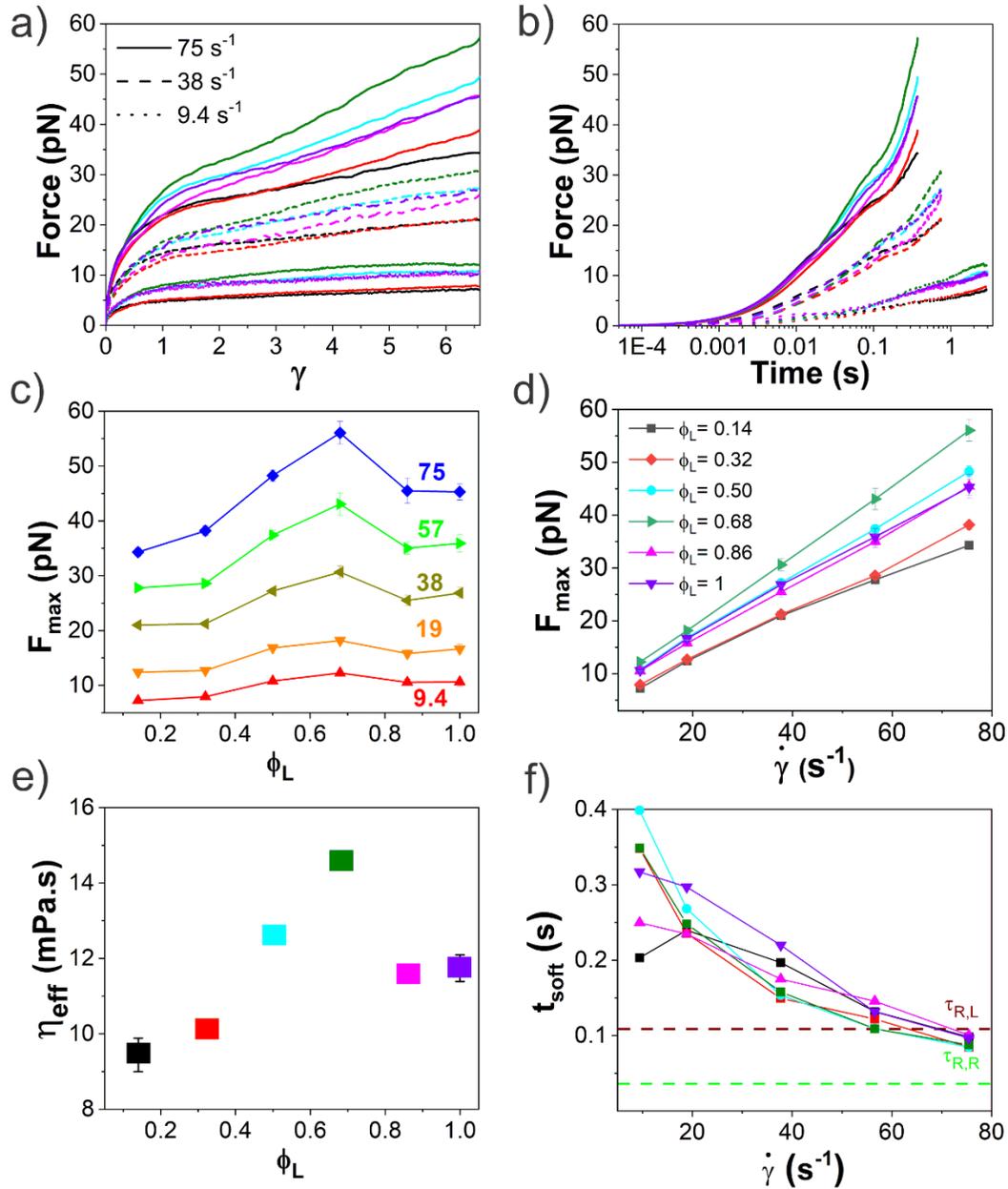

**Figure 3. Nonlinear force response of ring-linear blends exhibits a complex dependence on $\phi_L$ and strain rate.** (a) Measured force in response to nonlinear strain with rates of $\dot{\gamma}$=9.4, 38 and 75 s$^{-1}$ are shown. Data for all strain rates is shown in Fig. S1. Colors correspond to varying blend fractions as specified in the legend in (d). (b) Force responses from (a) plotted on a semi-log scale with respect to time. Strain hardening is evident at longer times and is more apparent at higher strain rates. (c) Maximum force values reached during strain as a function of $\phi_L$ for varying strain rates $\dot{\gamma}$ displayed in s$^{-1}$. (d) Maximum force values from (c) plotted as a function of strain rate $\dot{\gamma}$. (e) Effective viscosities $\eta_{eff}$ versus $\phi_L$ determined from the slopes of linear fits to the data shown in (d). (f) The softening time $t_{soft}$ as a function of $\dot{\gamma}$ for varying linear fractions as shown in legend. $t_{soft}$ shows a non-monotonic behavior with $\phi_L$ for lower strain rates but all values converge to the theoretical Rouse relaxation value for linear DNA at the highest strain rate. Theoretical values of Rouse relaxation for linear ($\tau_{R,L}$) and ring ($\tau_{R,R}$) polymers are shown as dashed lines.

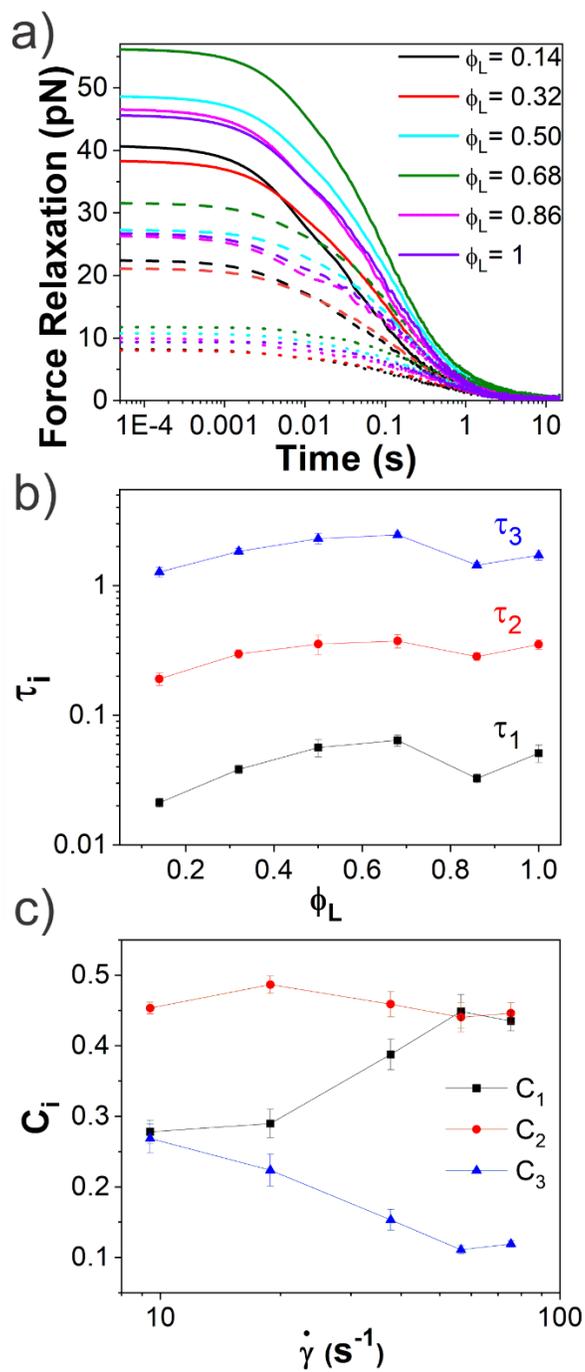

**Figure 4. Ring-linear DNA blends exhibit multi-mode relaxation following nonlinear strain**. (a) Force relaxation of all blends as a function of time following strain for strain rates $\dot{\gamma}$=9.4, 38 and 75 s$^{-1}$. Each curve is fit to a sum of three exponential decays with adjusted R-squared values of >0.99. (b) Time constants $\tau_1$, $\tau_2$ and $\tau_3$ determined from fits, averaged over all strain rates and plotted as a function of $\phi_L$. (c) Corresponding fractional amplitudes $C_1$, $C_2$ and $C_3$ determined from fits, averaged over all $\phi_L$ and plotted as a function of strain rate. Fractional amplitudes of $\tau_1$ and $\tau_3$ show a significant but opposite strain rate dependence whereas no strain rate dependence is observed for $\tau_2$.

**Viscoelastic properties of ring-linear DNA blends exhibit non-monotonic dependence on blend composition**

Karthik R. Peddireddy[1], Megan Lee[1], Charles M. Schroeder[2], Rae M. Robertson-Anderson[1,*]

[1]*Department of Physics and Biophysics, University of San Diego, 5998 Alcala Park, San Diego, CA 92110, United States*

[2]*Department of Materials Science and Engineering, Beckman Institute for Advanced Science and Technology & Department of Chemical and Biomolecular Engineering, University of Illinois at Urbana-Champaign, Urbana, IL 61801, United States*

**Supporting Information**

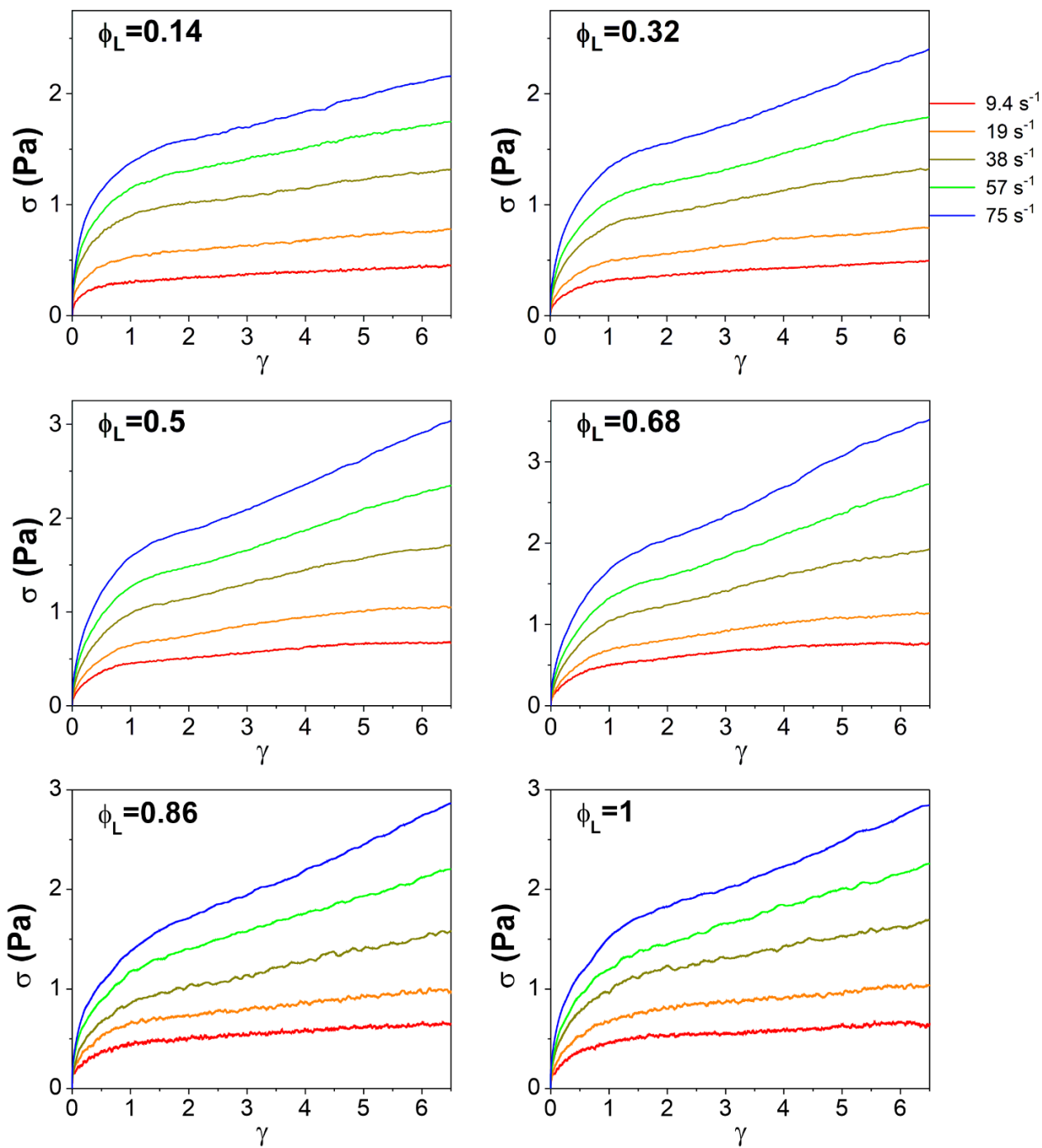

**Figure S1. Nonlinear stress responses of ring-linear DNA blends.** Nonlinear force ($F$) and stage displacement ($x$) data is converted into stress and strain via $\sigma = F/\pi R^2$ and $\gamma = x/2R$ where $R$ is the bead radius [1]. Linear fractions in each blend are shown as legends in each plot.

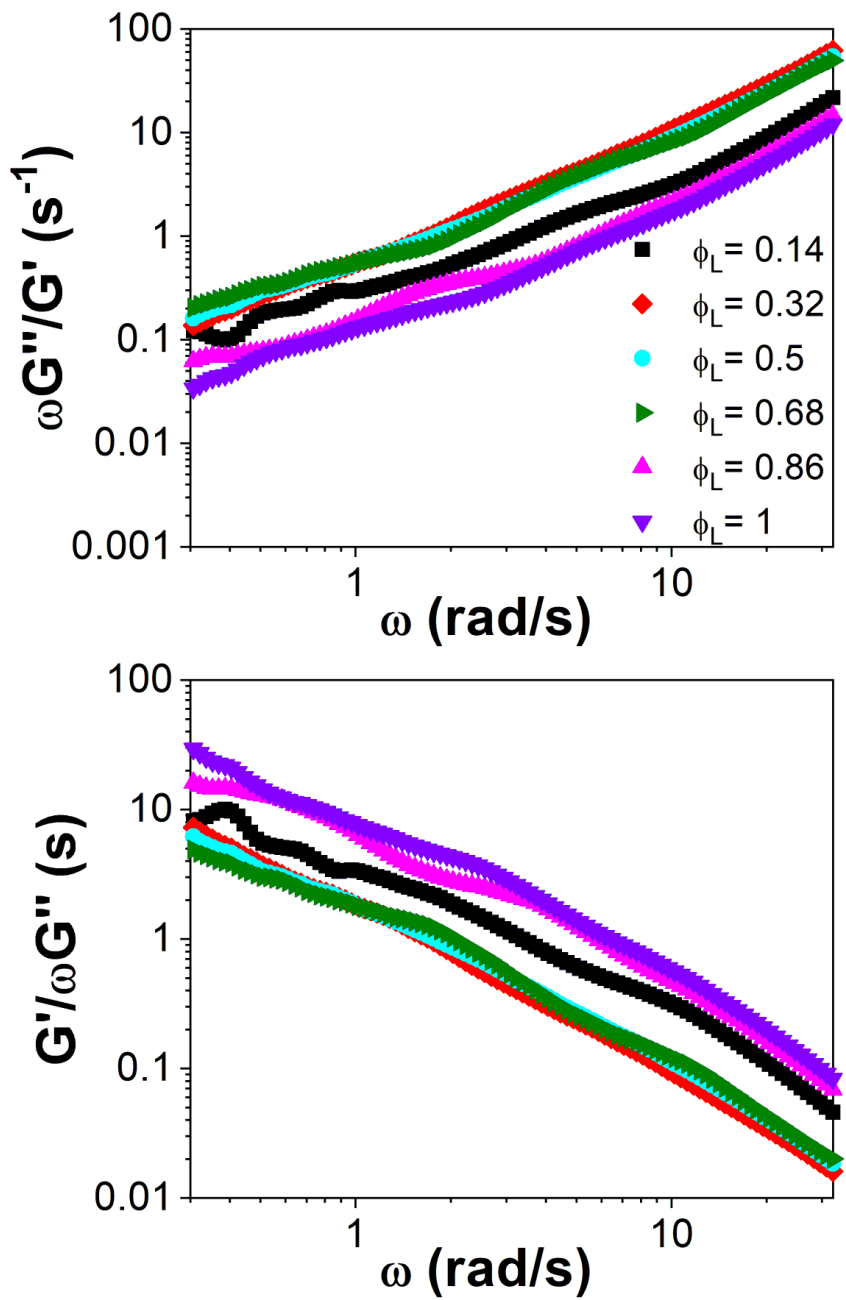

**Figure S2. Terminal relaxation frequency and time are determined from linear microrheology experiments.** (Top) Terminal relaxation frequencies, $\omega_T = \lim_{\omega \to 0} \omega G''/G'$, are clearly well below the frequency range ($\dot{\gamma}$=9.4-75s$^{-1}$) we explored for all blends. As shown in Fig. S1, applied nonlinear strain ($\gamma=\dot{\gamma}t$=6.7) far exceeds the critical value of 1. We are indeed probing the nonlinear regime as both necessary and sufficient conditions are met. (Bottom) Terminal relaxation times, $= \lim_{\omega \to 0} G'/\omega G''$, in the linear regime are at least an order of magnitude slower than the slowest relaxation time ($\tau_3$) measured in the nonlinear regime.

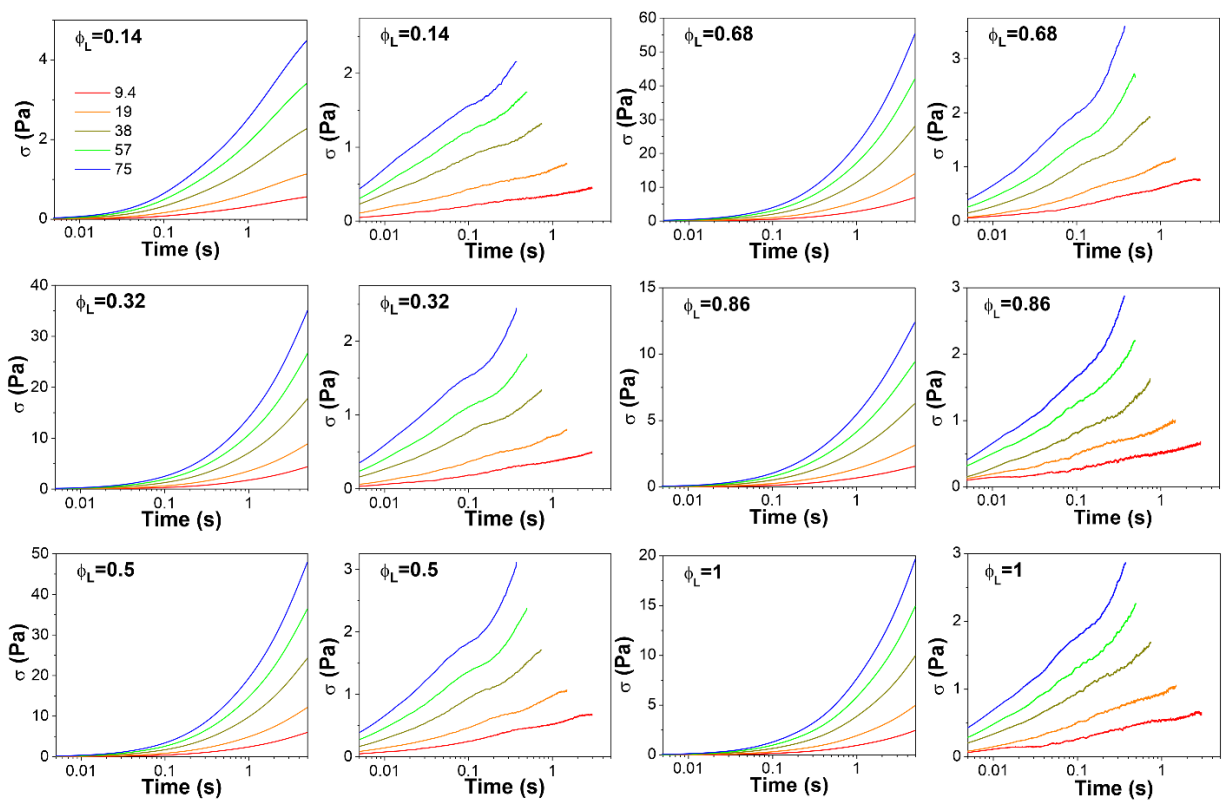

**Figure S3. Comparison between expected linear viscoelastic elastic (LVE) stress growth and measured nonlinear stress growth in nonlinear microrheology experiments.** (Odd columns) LVE stress growth is computed via $\sigma(t) = \dot{\gamma} \int_0^t G(t)\, dt$ [2]. Computed stress values strongly depend on $\phi_L$. Applied strain rates are shown in the top figure. (Even columns) Experimental nonlinear force response curves. Strain stiffening is increasingly observed with increasing strain rate in all blends.